\documentclass{aastex61}

\date{\today}

\begin{document}

\title{Planck 2015 constraints on the non-flat $\phi$CDM inflation model}

\author{Junpei Ooba}
\altaffiliation{ooba.jiyunpei@f.mbox.nagoya-u.ac.jp}
\affiliation{Department of Physics and Astrophysics, Nagoya University, Nagoya 464-8602, Japan}

\author{Bharat Ratra}
\affiliation{Department of Physics, Kansas State University, 116 Cardwell Hall, Manhattan, KS 66506, USA}

\author{Naoshi Sugiyama}
\affiliation{Department of Physics and Astrophysics, Nagoya University, Nagoya 464-8602, Japan}
\affiliation{Kobayashi-Maskawa Institute for the Origin of Particles and the Universe, Nagoya University, Nagoya, 464-8602, Japan}
\affiliation{Kavli Institute for the Physics and Mathematics of the Universe (Kavli IPMU), The University of Tokyo, Chiba 277-8582, Japan}



\begin{abstract}

We perform Markov chain Monte Carlo analyses to put constraints on the 
non-flat $\phi$CDM inflation model using Planck 2015 cosmic microwave 
background (CMB) anisotropy data and baryon acoustic oscillation distance
measurements. The $\phi$CDM model is a consistent dynamical dark energy
model in which the currently accelerating cosmological expansion is powered 
by a scalar field $\phi$ slowly rolling down an inverse power-law potential
energy density. We also use a physically consistent power spectrum for 
energy density inhomogeneities in this non-flat model. We find that, like 
the closed-$\Lambda$CDM and closed-XCDM models, the closed-$\phi$CDM model 
provides a better fit to the lower multipole region of the CMB temperature 
anisotropy data compared to that provided by the tilted flat-$\Lambda$CDM 
model. Also, like the other closed models, this model reduces the tension 
between the Planck and the weak lensing $\sigma_8$ constraints. However,
the higher multipole region of the CMB temperature anisotropy data are 
better fit by the tilted flat-$\Lambda$ model than by the closed models.
\end{abstract}




\section{Introduction} \label{sec:intro}

The standard cosmological model, spatially-flat $\Lambda$CDM 
\citep{Peebles1984}, is described by six cosmological parameters:
$\Omega_{\rm b} h^2$ and $\Omega_{\rm c} h^2$, the current values
of the baryonic and cold dark matter (CDM) density parameters
multiplied by the square of the Hubble constant $H_0$ (in units of 
100 km s$^{-1}$  Mpc$^{-1}$); $\theta$, the angular diameter distance as 
a multiple of the 
sound horizon at recombination; $\tau$, the reionization optical depth; 
and $A_{\rm s}$ and $n_{\rm s}$, the amplitude and spectral index of the
(assumed) power-law primordial scalar energy density inhomogeneity power 
spectrum, \citep{Adeetal2016a}. In this model, the currently accelerated 
cosmological expansion is powered by the cosmological constant $\Lambda$,
which is equivalent to a dark energy ideal fluid with equation of state 
parameter $w_0 = -1$. This model assumes flat spatial hypersurfaces, 
which is largely consistent with most available observational constraints 
\citep[][and references therein]{Adeetal2016a}. 

However, using a physically consistent non-flat inflation model power 
spectrum of energy density inhomogeneities \citep{RatraPeebles1995, Ratra2017},
we recently found that cosmic microwave background (CMB) anisotropy 
measurements do not require flat spatial hypersurfaces in the six 
parameter non-flat $\Lambda$CDM
model\footnote{Here, compared to the flat-$\Lambda$CDM model, $n_s$ is 
replaced by the current value of the spatial curvature density parameter 
$\Omega_{\rm k}$.} \citep{Oobaetal2018a, ParkRatra2018a, ParkRatra2018b} and 
also in the seven parameter XCDM 
model \citep{Oobaetal2017, ParkRatra2018b}, in which the equation of state 
relating the
pressure and energy density of the dark energy fluid is written as 
$p_X = w_0 \rho_X$ and $w_0$ is the additional, seventh, 
parameter.\footnote{Again, $n_s$ is replaced by $\Omega_{\rm k}$.}
Also, there are suggestions that flat-$\Lambda$CDM might not be 
as compatible with more recent, larger compilations of measurements  
\citep{Solaetal2017a, Solaetal2017b, Solaetal2018, Solaetal2017c, Zhangetal2017}
that might be more consistent with dynamical dark energy models.
These include the simplest, physically consistent, seven parameter
spatially-flat $\phi$CDM model in which a scalar field $\phi$ with 
potential energy density $V(\phi) \propto \phi^{-\alpha}$ is the dynamical 
dark energy \citep{PeeblesRatra1988, RatraPeebles1988} and $\alpha > 0$ 
is the seventh parameter that governs dark energy evolution.\footnote{While
XCDM is often used to model dynamical dark energy, it is not a physically 
consistent model as it cannot describe the the evolution of energy density 
inhomogeneities. Also, XCDM does not accurately model $\phi$CDM dark energy
dynamics \citep{PodariuRatra2001}.} 

Non-zero spatial curvature brings in a new length scale, in addition to 
the Hubble scale. Consequently, in non-flat models it is incorrect to assume a 
power law spectrum for energy density inhomogeneities. Instead in the non-flat
case one must use an inflation model to compute a consistent power spectrum.
For open spatial hypersurfaces the \citet{Gott1982} open-bubble inflation 
model is taken as the initial epoch of the cosmological model and one 
computes zero-point quantum fluctuations during the open inflation epoch and 
propagates these to the current open accelerating universe where they are 
energy density inhomogeneities with a power spectrum that is not a 
power law \citep{RatraPeebles1994, RatraPeebles1995}.\footnote{For discussions 
of observational consequences of the open inflation model see \citet{Kamionkowskietal1994}, \citet{Gorskietal1995}, \citet{Gorskietal1998}, and 
references therein.}
For closed spatial hypersurfaces Hawking's prescription for the quantum state 
of the universe \citep{Hawking1984} can be used to construct a closed 
inflation model \citep{Ratra1985, Ratra2017}.
Zero-point quantum-mechanical fluctuations during closed inflation 
provide a late-time energy density inhomogeneity power spectrum that is not 
a power law \citep{Ratra2017}; it is a generalization to the 
closed case \citep{WhiteScott1996, Starobinsky1996, Zaldarriagaetal1998, Lewisetal2000, LesgourguesTram2014} of the flat-space scale-invariant 
spectrum.

Compared to the six parameter flat-$\Lambda$CDM inflation model discussed 
above, in the non-flat case there is no simple tilt option, so 
$n_{\rm s}$ is no longer a free parameter and is replaced by
$\Omega_{\rm k}$ which results in 
the six parameter non-flat $\Lambda$CDM model \citep{Oobaetal2018a}
or the seven parameter non-flat XCDM model \citep{Oobaetal2017}.
Here, we use a physically consistent non-flat seven parameter $\phi$CDM 
scalar field dynamical dark energy model \citep{Pavlovetal2013}, again 
with $n_s$ replaced by $\Omega_{\rm k}$.\footnote{We study the seven parameter
flat $\phi$CDM model elsewhere \citep{Oobaetal2018b}.}   
In this paper we use the Planck 2015 CMB anisotropy data to constrain this 
seven parameter non-flat $\phi$CDM inflation model. We find in this model that 
the Planck 2015 CMB anisotropy data in conjunction with baryon acoustic 
oscillation (BAO) measurements do not require that spatial hypersurfaces 
be flat. The data favor a mildly closed model. These results are consistent 
with those from our earlier analyses of the six parameter non-flat $\Lambda$CDM 
inflation model \citep{Oobaetal2018a} and the seven parameter non-flat
XCDM inflation model \citep{Oobaetal2017}.

  The structure of our paper is as follows. In Sec.\ II we summarize the methods
we use in our analyses here. Our parameter constraints are tabulated, plotted, 
and discussed in Sec.\ III, where we also attempt to judge how well the 
best-fit closed-$\phi$CDM model fits the data. We conclude in Sec.\ IV.
\\


\begin{figure}[ht]
\plotone{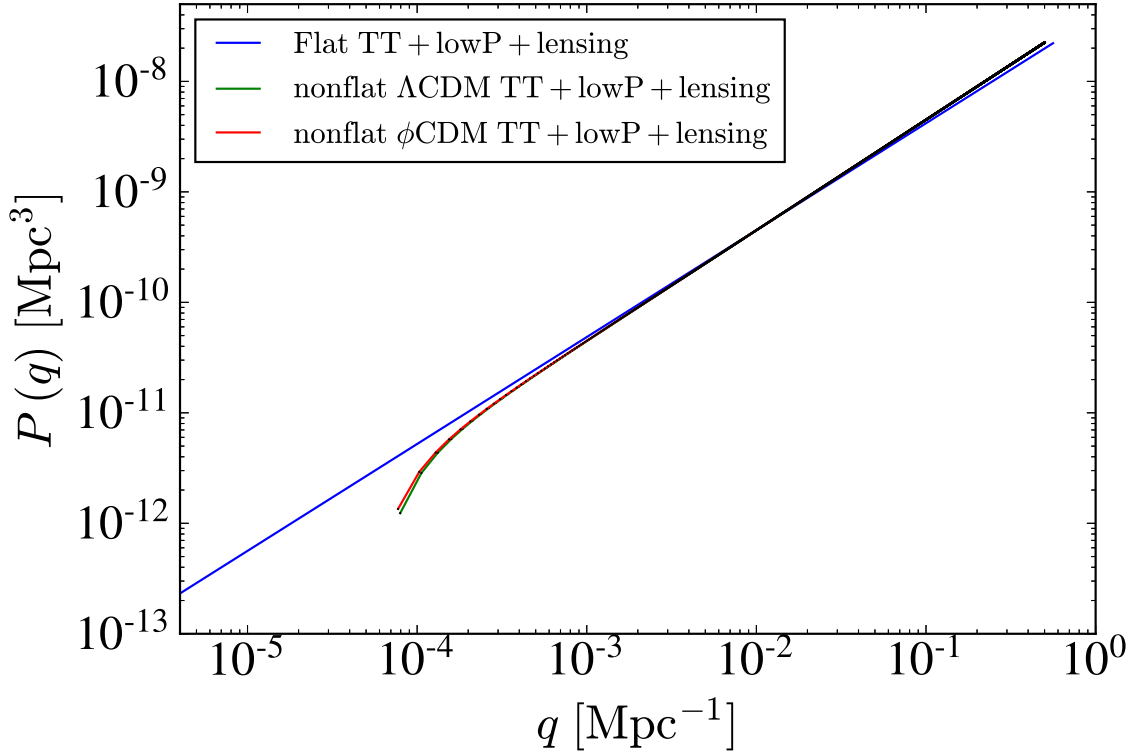}
\caption{Best-fit (see text) gauge-invariant fractional energy density 
inhomogeneity power spectra. The blue line corresponds to the tilted
flat-$\Lambda$CDM model of \citet{Adeetal2016a}. In the closed case, 
wavenumber $q \propto A + 1$ where the eigenvalue 
of the spatial Laplacian is $-A(A+2)$, $A$ is a non-negative integer, with 
$A = 0$ corresponding to the constant zero-mode on the three sphere, the power 
spectrum vanishes at $A = 1$, and the points on the red and green curves 
correspond to $A = 2, 3, 4, ...$, see eqns.\ (8) and (203) of 
\citet{Ratra2017}. On large scales the power spectra for the best-fit 
closed $\phi$CDM (red curve) and closed $\Lambda$CDM (green curve) models are
suppressed relative to that of the best-fit tilted flat-$\Lambda$CDM model.
$P(q)$ is normalized by the best-fit value of $A_{\rm s}$ at the pivot scale 
$k_0 = 0.05$ [Mpc${}^{-1}$]. 
\label{fig:Pk}}
\end{figure}

\section{Methods}

The equations of motion of the non-flat $\phi$CDM model \citep{Pavlovetal2013} 
are
\begin{eqnarray}
\label{eq:eom}
& \ddot\phi + 3 {\dot a \over a} \dot\phi - \kappa \alpha m_P^2 
  \phi^{-(\alpha + 1)} = 0, \\
& \left({\dot a \over a}\right)^2 = {8 \pi \over 3 m_P^2}( \rho + \rho_\phi) 
  - {k \over a^2}, \\
& \rho_\phi = {m_P^2 \over 32 \pi} \left( \dot\phi^2 + 2 \kappa m_P^2  
  \phi^{-\alpha}\right) .
\end{eqnarray}
Here $\phi$ is the dark energy scalar field with potential energy density 
$V(\phi) = \kappa m_P^2 \phi^{-\alpha}$, $\alpha > 0$, $m_p$ is the Planck 
mass, and the parameter $\kappa$ is determined in terms of the other 
parameters. Also, $a$ is the scale factor, $k$ is the curvature parameter 
that takes values -1, 0, 1, an overdot denotes a time derivative, $\rho$ is 
the energy density of baryonic and cold matter as well as radiation and 
neutrinos, and $\rho_\phi$ that of the scalar field. The $\phi$CDM model 
equations of motion has a time-dependent attractor or
tracker solution \citep{PeeblesRatra1988,RatraPeebles1988,Pavlovetal2013}.   

We use the open and closed inflation model quantum energy density 
inhomogeneity power spectrum \citep{RatraPeebles1995, Ratra2017} in our 
analyses of the non-flat $\phi$CDM model.
Figure \ref{fig:Pk} shows best-fit power spectra for the non-flat $\phi$CDM and 
non-flat $\Lambda$CDM inflation models as well as a tilted flat-$\Lambda$CDM 
inflation model power spectrum.
In this study we compute the angular power spectra of the CMB anisotropy by 
using CLASS \citep{Blasetal2011}\footnote{Our flat space $\phi$CDM CMB 
anisotropy angular power spectra differ somewhat from earlier results in 
\citet{Braxetal2000} and \citet{Mukherjeeetal2003}. We have verified that 
our results are accurate.}
and perform the Markov chain Monte Carlo analyses with Monte Python \citep{Audrenetal2013}.

The ranges of the cosmological parameters we consider are
\begin{eqnarray}
\label{eq:prior}
&100\theta \in (0.5,10),\ \ \Omega_{\rm b}h^2 \in (0.005,0.04),\ \ \Omega_{\rm c}h^2 \in (0.01,0.5), \nonumber\\
&\tau \in (0.005,0.5),\ \ {\rm ln}(10^{10}A_{\rm s}) \in (0.5,10),\ \ \Omega_{\rm k} \in (-0.5, 0.5),\ \ \alpha \in (0, 8).
\end{eqnarray}
The CMB temperature and the effective number of neutrinos were set to 
$T_{\rm CMB}= 2.7255\ \rm K$ from COBE \citep{Fixsen2009} and $N_{\rm eff}=3.046$ 
with one massive (0.06 eV) and two massless neutrino species in a normal 
hierarchy. The primordial helium fraction $Y_{\rm He}$ is 
inferred from standard Big Bang nucleosynthesis, as a function of the baryon 
density.

We constrain model parameters by comparing our results to the CMB angular 
power spectrum data from the
Planck 2015 mission \citep{Adeetal2016a} and the BAO measurements from the 
matter power spectra obtained by the 6dF Galaxy Survey (6dFGS) 
\citep{Beutleretal2011}, the Baryon Oscillation Spectroscopic Survey 
(BOSS; LOWZ and CMASS) \citep{Andersonetal2014}, and the Sloan Digital Sky 
Survey (SDSS) main galaxy sample (MGS) \citep{Rossetal2015}.

\section{Results}

\begin{figure}[ht]
\plottwo{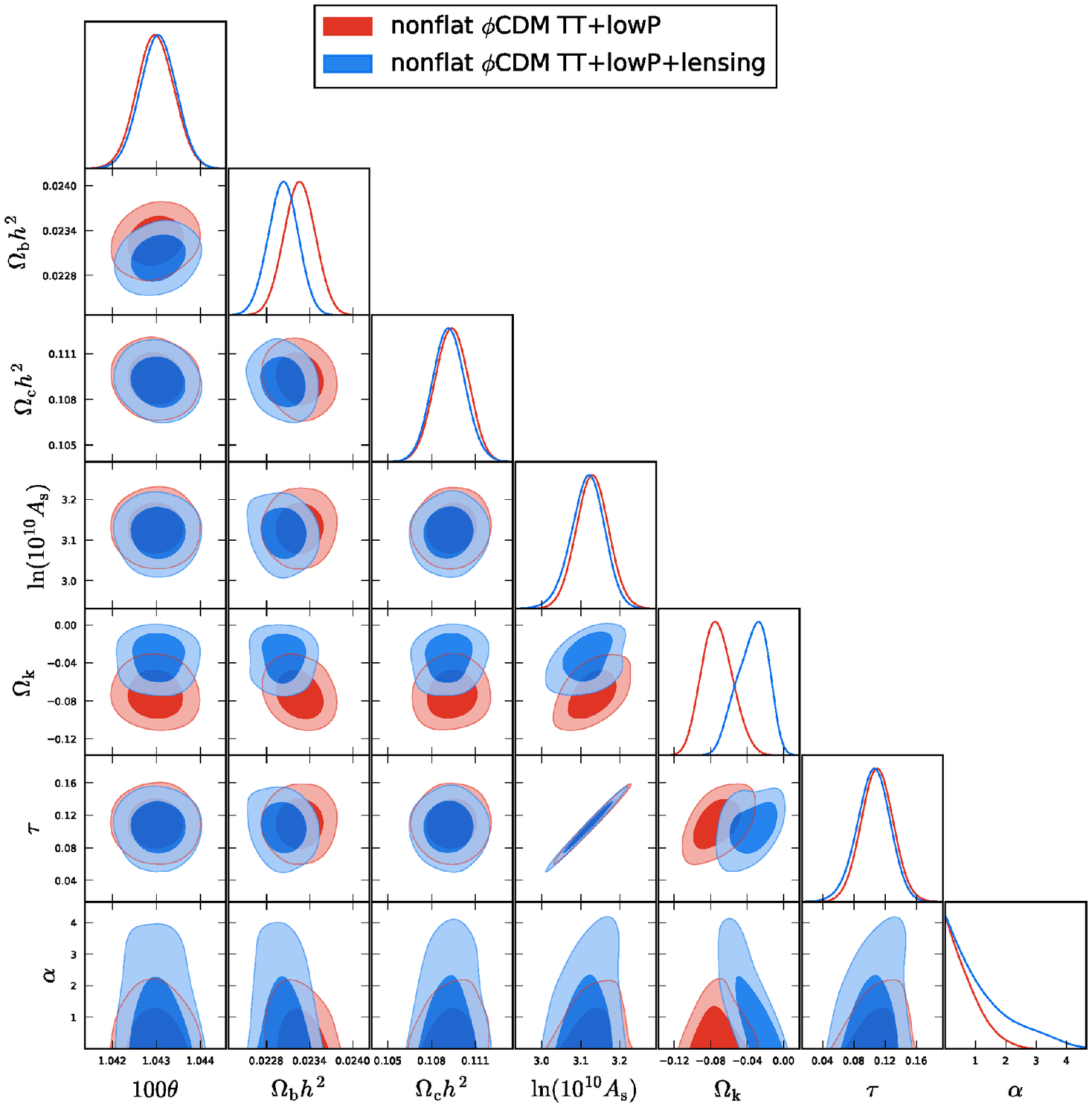}{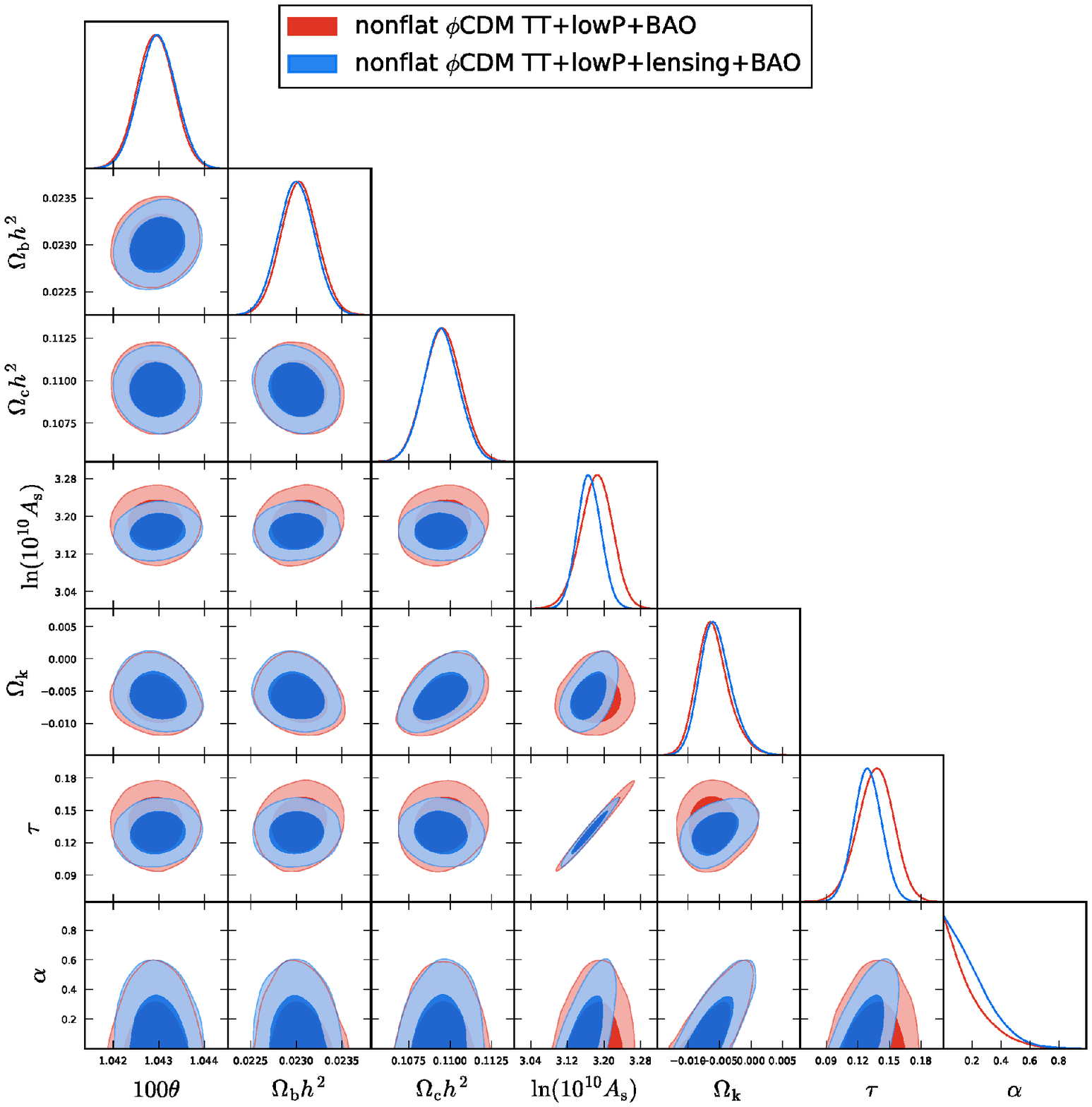}
\caption{$68.27\%$ and $95.45\%$ confidence level contours for the non-flat 
$\phi$CDM inflation model using various data sets, with the other parameters 
marginalized.\label{fig:tri}}
\end{figure}

\begin{figure}[ht]
\gridline{\fig{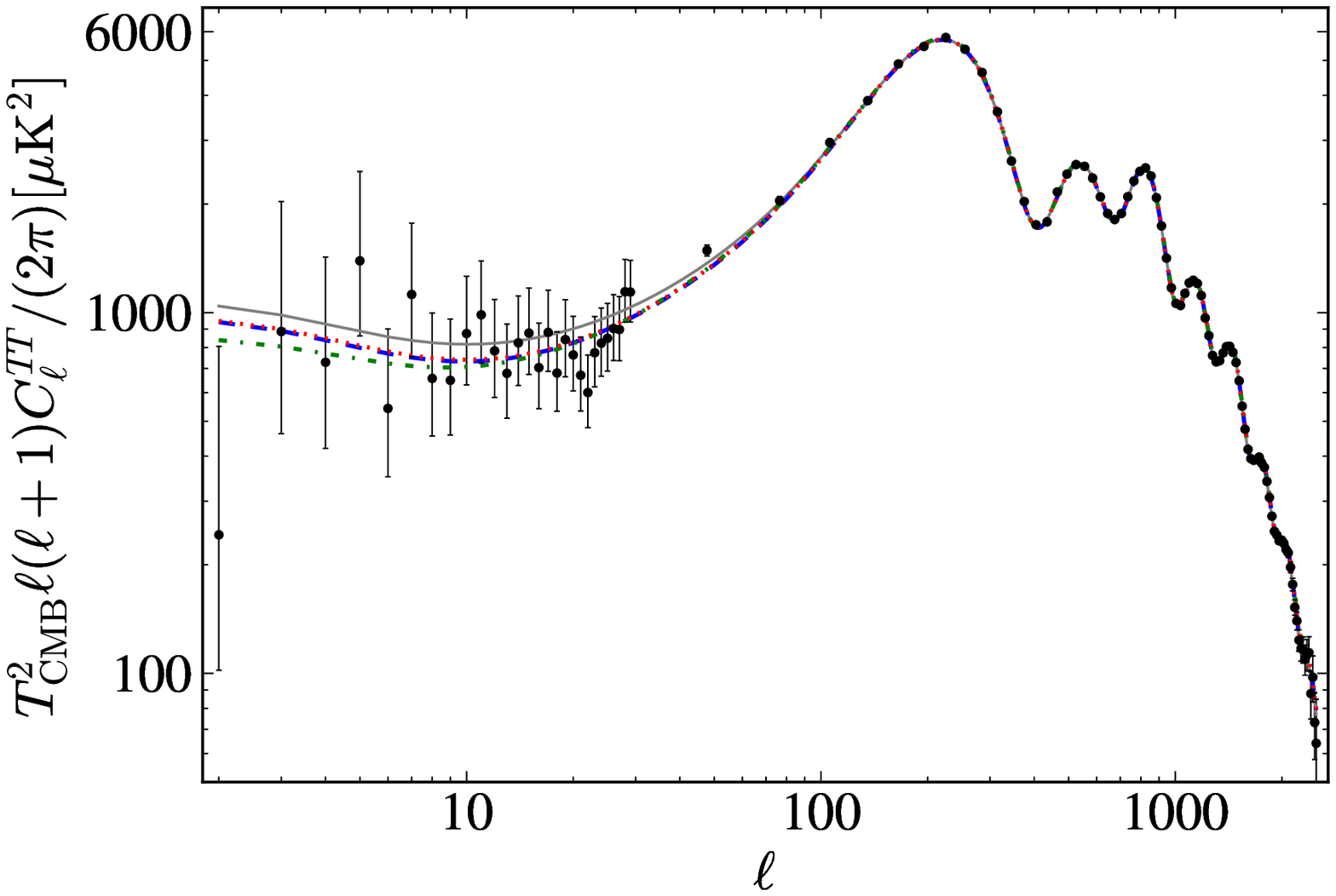}{0.45\textwidth}{(a)}
          \fig{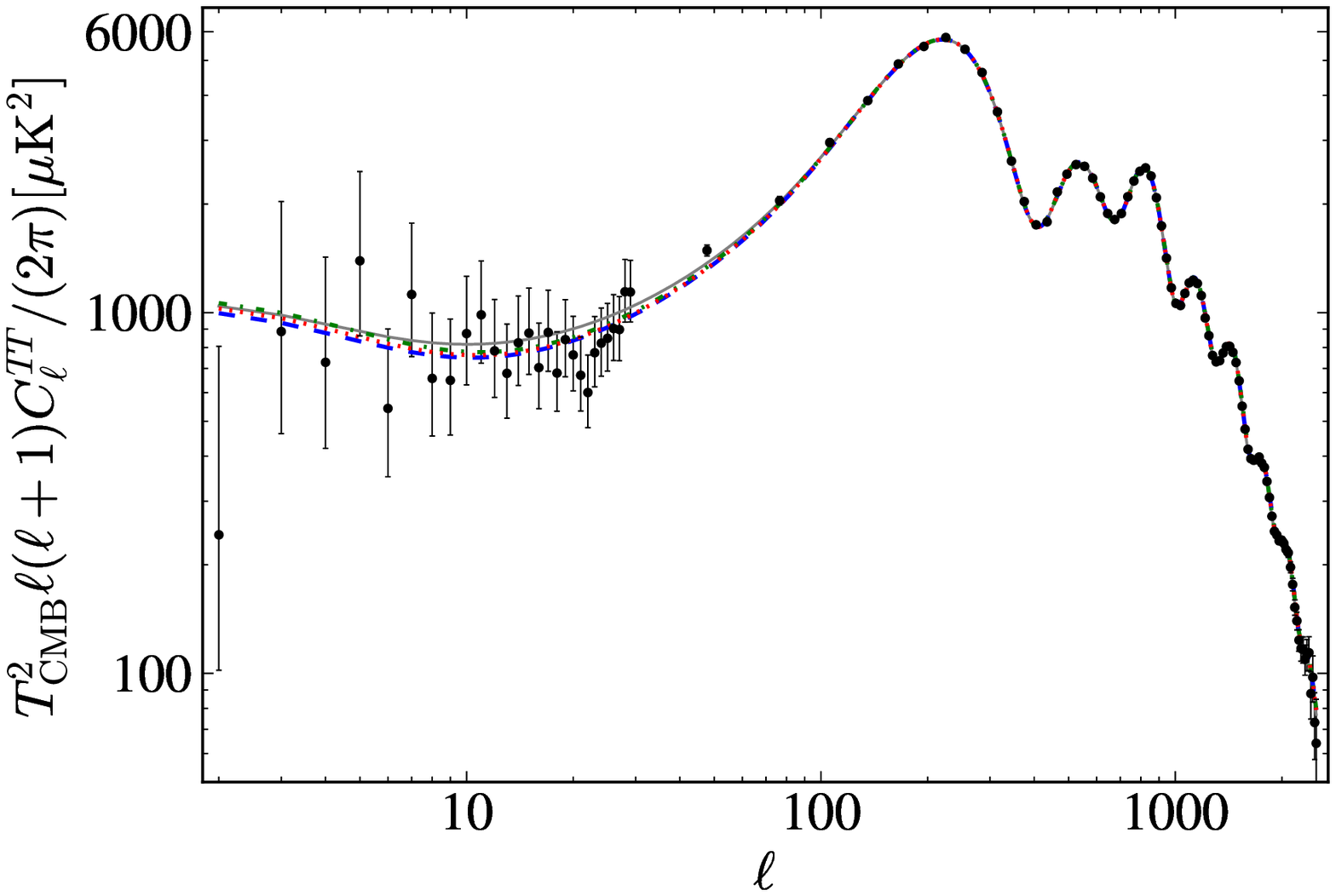}{0.45\textwidth}{(b)}
          }
\gridline{\fig{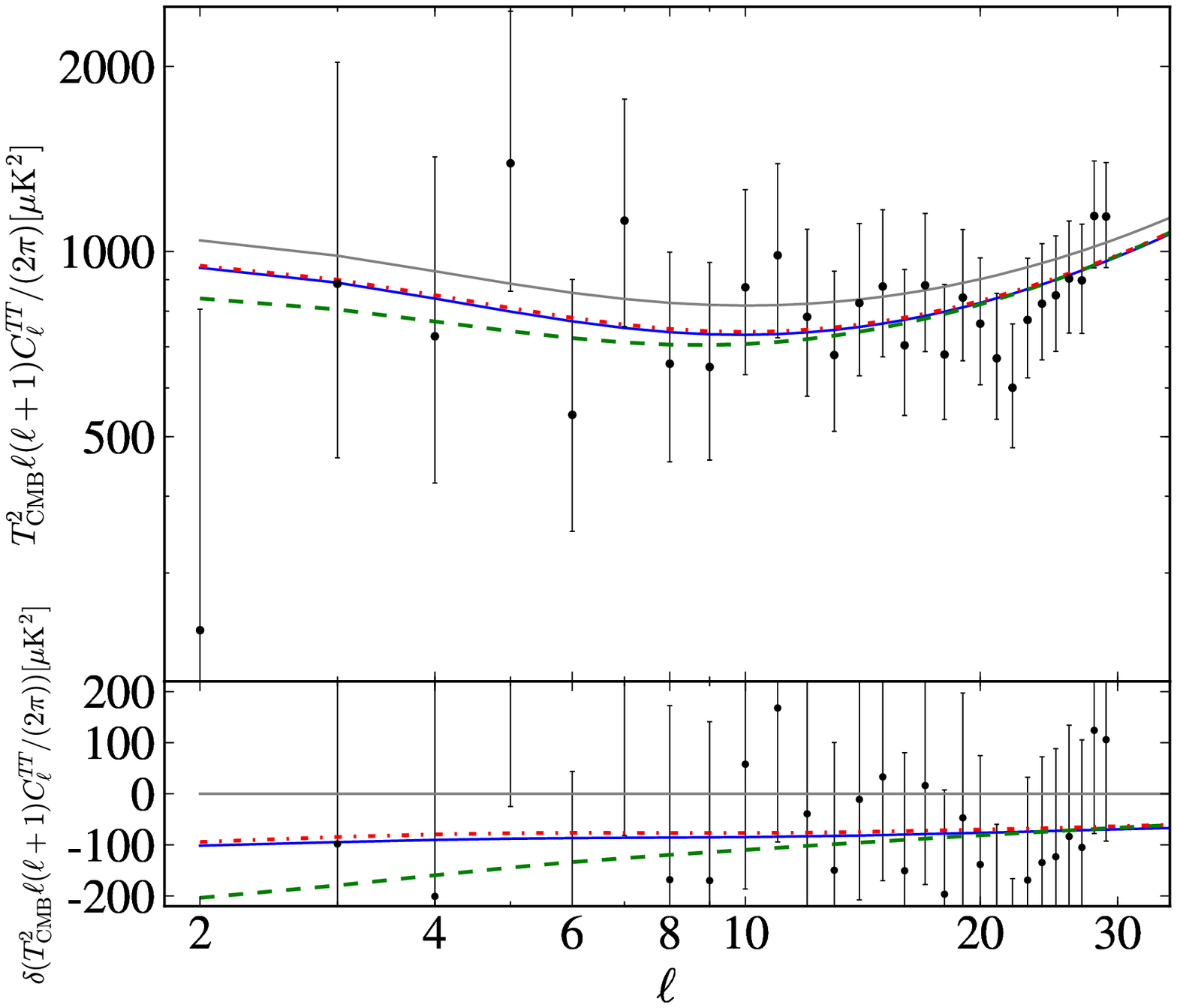}{0.42\textwidth}{(c)}
          \fig{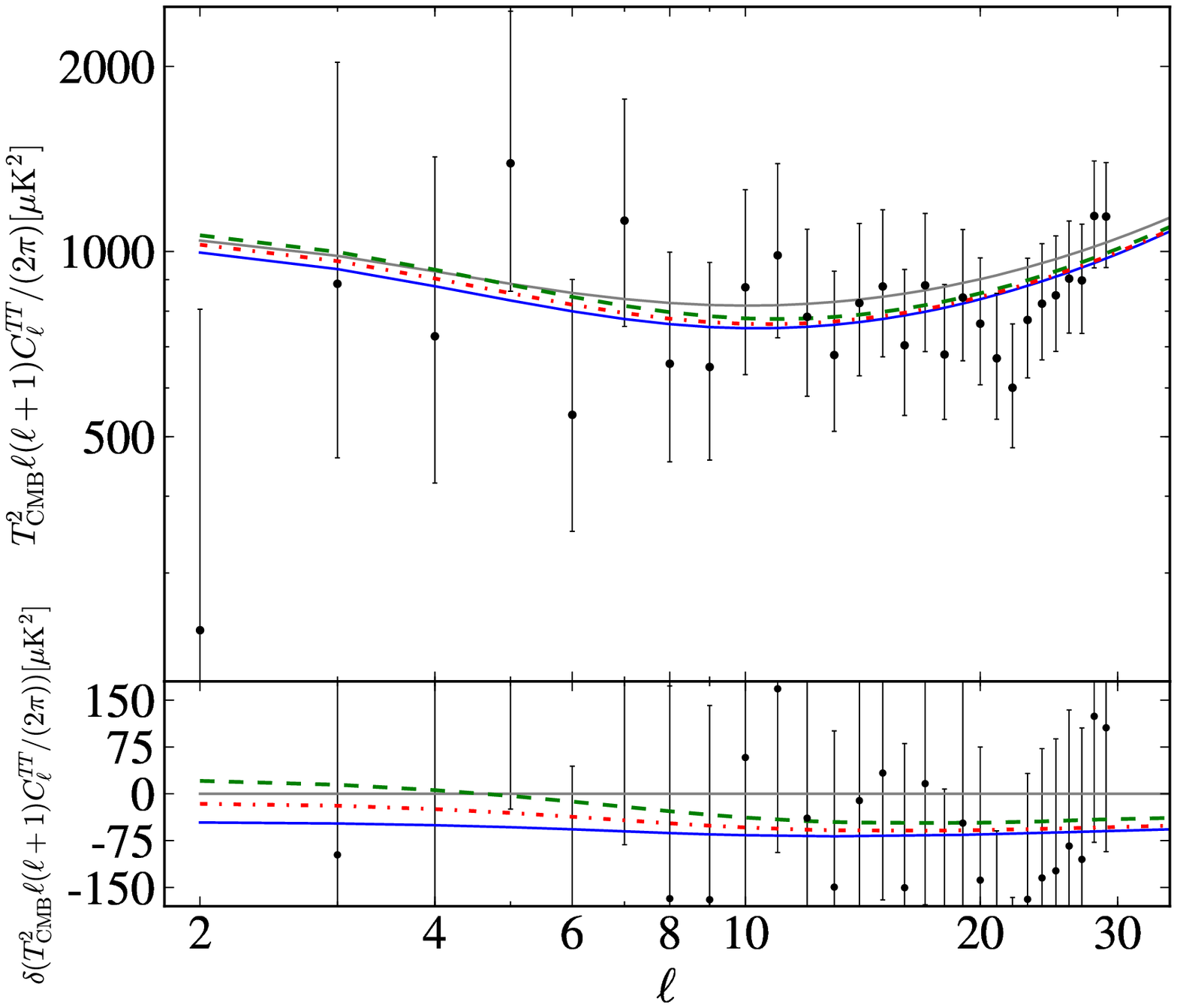}{0.42\textwidth}{(d)}
          }
\gridline{\fig{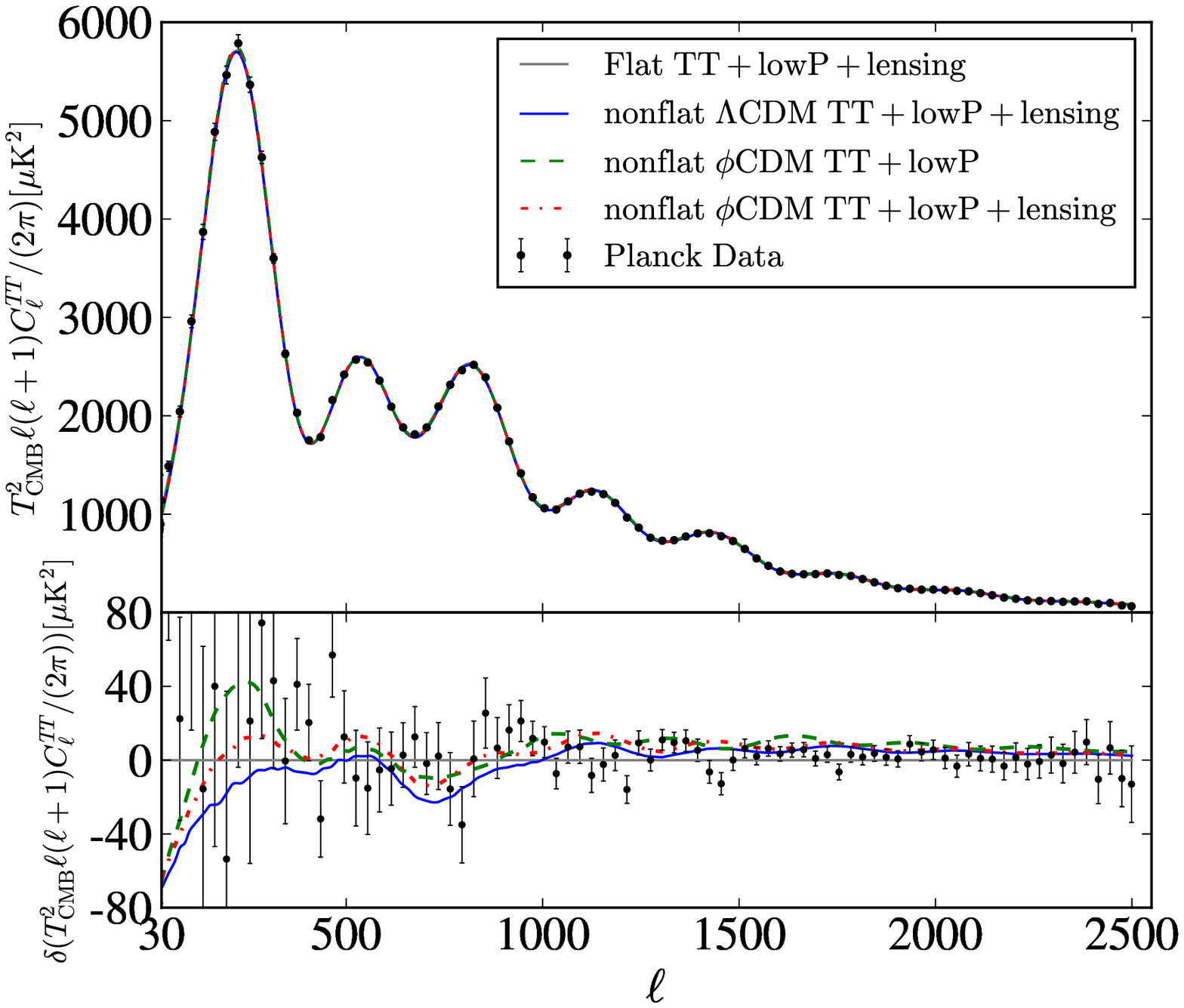}{0.43\textwidth}{(e)}
          \fig{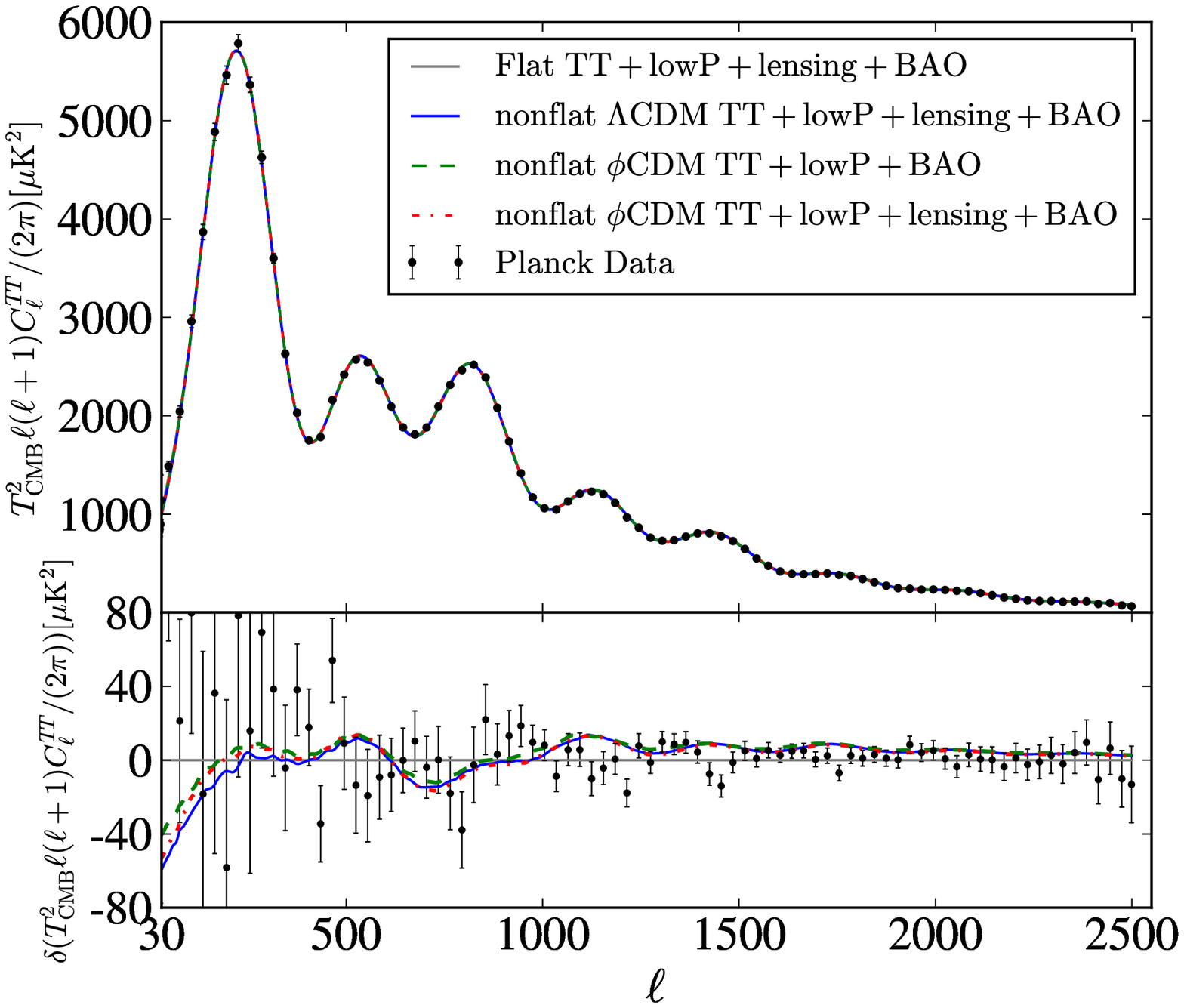}{0.43\textwidth}{(f)}
          }
\caption{The $C_{\ell}$ for the best-fit non-flat $\phi$CDM, non-flat $\Lambda$CDM 
and spatially-flat tilted $\Lambda$CDM (gray solid line) models. Linestyle 
information are in the boxes in the two lowest panels. Planck 2015 data are 
shown as black points with error bars. Left panels (a), (c) and (e) are from 
CMB data alone analyses, while right panels (b), (d) and (f) analyses also
include BAO data. The top panels show the all-$\ell$ region. The middle 
panels show the low-$\ell$ region $C_\ell$ and residuals. The bottom panels show the 
high-$\ell$ region $C_\ell$ and residuals.\label{fig:cls}}
\end{figure}

\begin{figure}[ht]
\plottwo{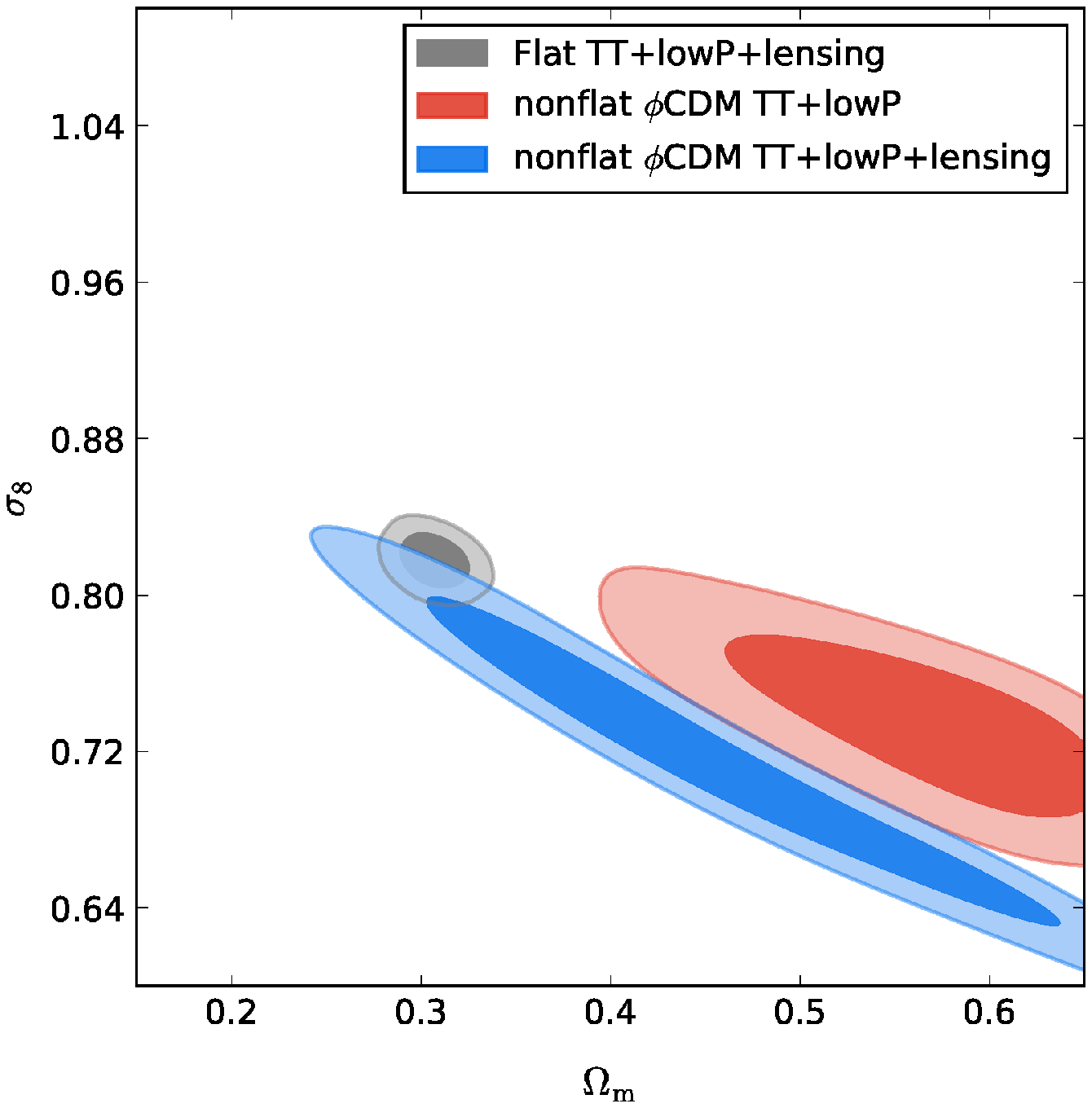}{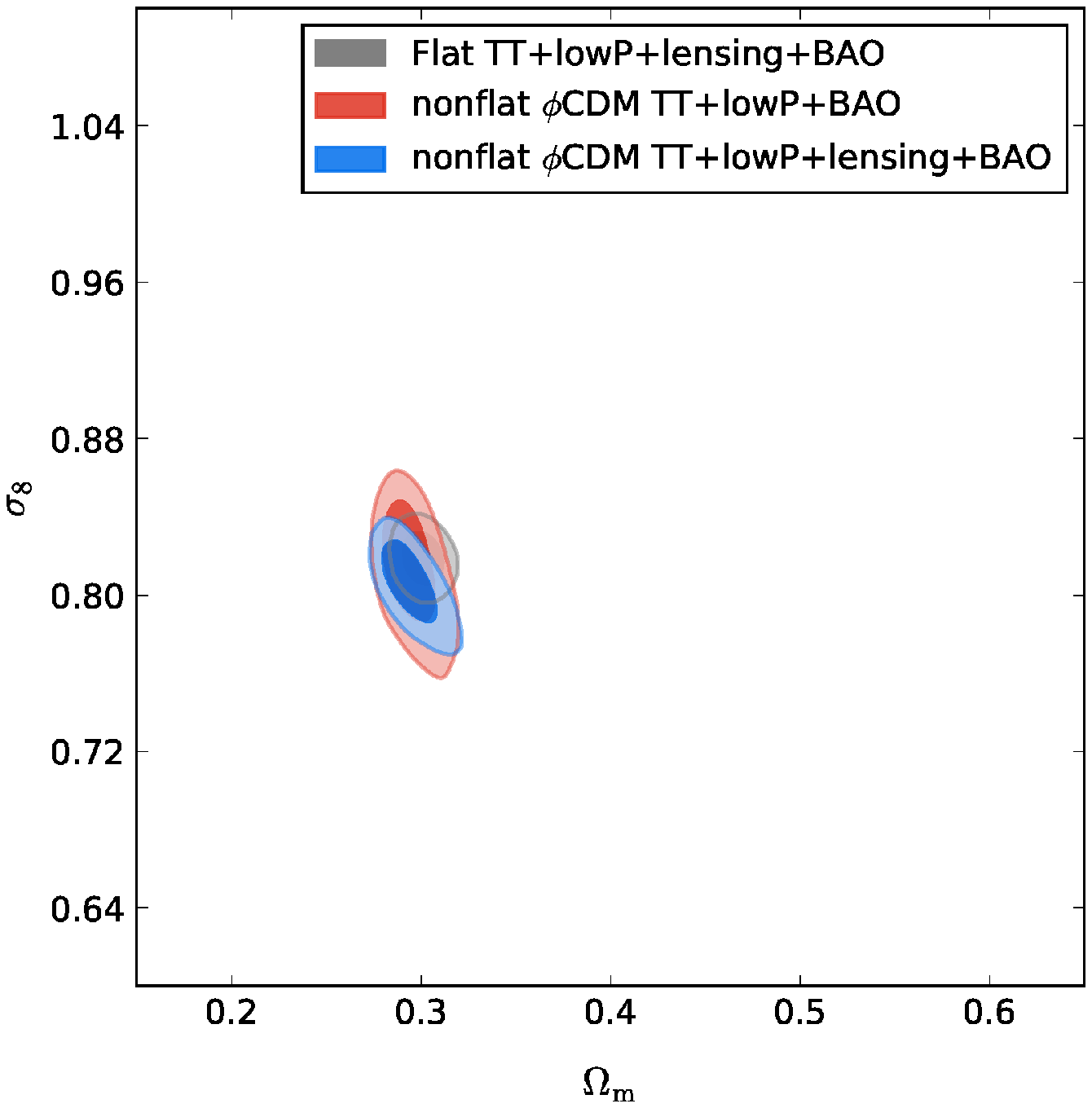}
\caption{$68.27\%$ and $95.45\%$ confidence level contours in 
the $\sigma_8$--$\Omega_{\rm m}$ plane.\label{fig:sigm}}
\end{figure}

In this section we tabulate, plot, and discuss the resulting constraints 
on the seven parameter non-flat $\phi$CDM inflation model. 
Table \ref{tab:table1} lists mean values and $68.27\%$ limits on the 
cosmological parameters ($95.45\%$ upper limits on $\alpha$),
and Fig. \ref{fig:tri} shows two-dimensional constraint contours and 
one-dimensional likelihoods from the 4 different CMB and BAO data sets 
used in this study. Here all other parameters are marginalized.
We set an additional prior on the Hubble constant, $h\geq0.45$, for the CMB data only cases to realize convergence in a reasonable amount of 
time.\footnote{We thank C.-G.\ Park for discussions about this and for pointing out a numerical error in our initial analyses. Our corrected 
results here are in very good agreement with those of \citet{ParkRatra2018c}.}  
CMB temperature anisotropy spectra for the best-fit non-flat $\phi$CDM models are shown in Fig. \ref{fig:cls},
compared with non-flat and tilted spatially-flat $\Lambda$CDM models.
Contours at $68.27\%$ and $95.45\%$ confidence level in the $\sigma_8$--$\Omega_{\rm m}$ plane
are shown in Fig. \ref{fig:sigm} with other parameters marginalized.

\begin{table*}[ht]
\caption{\label{tab:table1}
68.27\% (or 95.45\% on $\alpha$) confidence limits on cosmological parameters of the non-flat $\phi$CDM model from CMB and BAO data.}
\centering
\begin{tabular}{lcccc}
\hline
\hline
\textrm{Parameter}&
\textrm{TT+lowP ($h\geq0.45$)}&
\textrm{TT+lowP+lensing ($h\geq0.45$)}&
\textrm{TT+lowP+BAO}&
\textrm{TT+lowP+lensing+BAO}\\
\hline
$\Omega_{\rm b}h^2$ & $0.02326\pm 0.00022$ & $0.02303\pm 0.00020$ & $0.02303\pm 0.00020$ & $0.02300\pm 0.00020$\\
$\Omega_{\rm c}h^2$ & $0.1094\pm 0.0011$ & $0.1091\pm 0.0011$ & $0.1095\pm 0.0011$ & $0.1095\pm 0.0011$\\
$100\theta$ & $1.04296\pm 0.00041$ & $1.04303\pm 0.00041$ & $1.04293\pm 0.00041$ & $1.04298\pm 0.00041$\\
$\tau$ & $0.110\pm 0.020$ & $0.104\pm 0.021$ & $0.137\pm 0.018$ & $0.129\pm 0.013$\\
${\rm ln}(10^{10}A_{\rm s})$ & $3.130\pm 0.041$ & $3.116\pm 0.042$ & $3.183\pm 0.036$ & $3.168\pm 0.026$\\
$\Omega_{\rm k}$ & $-0.073\pm 0.017$ & $-0.034\pm 0.016$ & $-0.006\pm 0.003$ & $-0.006\pm 0.003$\\
$\alpha$\ $[2\sigma\ {\rm limit}]$ & $<1.82$ & $<3.26$ & $<0.305$ & $<0.304$\\
\hline
$H_0$ [km/s/Mpc] & $48.76\pm 2.39$ & $54.70\pm 6.86$ & $69.13\pm 1.06$ & $67.09\pm 1.11$\\
$\Omega_{\rm m}$ & $0.56\pm 0.06$ & $0.46\pm 0.12$ & $0.29\pm 0.01$ & $0.29\pm 0.01$\\
$\sigma_8$ & $0.735\pm 0.031$ & $0.716\pm 0.053$ & $0.815\pm 0.021$ & $0.806\pm 0.014$\\
\hline
\hline
\end{tabular}
\end{table*}

We find that the $95.45\%$ limits of the spatial curvature density parameter are
\begin{eqnarray}
\label{eq:omk}
\Omega_{\rm k} &=& -0.034 {+0.028 \atop -0.033} \ \ (95.45\%,\ \rm TT+lowP+lensing),\\
\Omega_{\rm k} &=& -0.006 \pm 0.005\ \ (95.45\%,\ \rm TT+lowP+lensing+BAO).
\end{eqnarray}
Both data sets result in best-fit models about 3$\sigma$ away from flat.
We note that many analyses based on a variety of 
different, non-CMB, observations also do not rule out non-flat dark energy 
models \citep{Farooqetal2015, Saponeetal2014, Lietal2014, Caietal2016, Chenetal2016, YuWang2016, LHuillierShafieloo2017, Farooqetal2017, Lietal2016, WeiWu2017, Ranaetal2017, Yuetal2018, Mitraetal2018, Ryanetal2018}.

From the TT + lowP + lensing + BAO column in Table \ref{tab:table1}, we see that
$H_0 = 67.09 \pm 1.11$ km s${}^{-1}$ Mpc${}^{-1}$.  This is consistent with
most other determinations \citep{ChenRatra2011, Calabreseetal2012, Sieversetal2013, Aubourgetal2015, LHuillierShafieloo2017, Lukovicetal2016, Chenetal2017, Wangetal2017, LinIshak2017, Abbottetal2017, Yuetal2018}, but a little lower 
than local determinations \citep{Freedmanetal2012, Riessetal2016}. 
The $\Omega_{\rm m}$ result is also consistent with most other determinations
\citep[e.g.,][]{ChenRatra2003}.

All three closed inflation models are more 
consistent with the low-$\ell$ $C_{\ell}$ observations\footnote{At low $\ell$, polarization anisotropy systematics (and possibly foreground contamination) are more significant, as is cosmic variance. Our analyses here assume that the low-$\ell$ error bars account for all relevant effects.}
 and the weak lensing 
$\sigma_8$ constraints than is the best fit spatially-flat tilted $\Lambda$CDM,
but they do worse at fitting the higher-$\ell$ ${C_{\ell}}$ measurements.

\begin{table*}[ht]
\caption{\label{tab:table2}
Minimum $\chi^2_{\rm eff}$ values for the best-fit closed-$\phi$CDM (and tilted flat-$\Lambda$CDM)
inflation model.}
\centering
\begin{tabular}{lcc}
\hline
\hline
\textrm{Data sets}& \textrm{$\chi^2_{\rm eff}$}& \textrm{d.o.f.}\\
\hline
TT+lowP ($h\geq0.45$) & 11272 (11262) & 188 (189)\\
TT+lowP+lensing ($h\geq0.45$) & 11294 (11272) & 196 (197)\\
TT+lowP+BAO & 11287 (11266) & 192 (193)\\
TT+lowP+lensing+BAO & 11298 (11277) & 200 (201)\\
\hline
\hline
\end{tabular}
\end{table*}

\begin{table*}[ht]
\caption{\label{tab:table3}
$\chi^2$ values for the best-fit closed-$\phi$CDM (and tilted flat-$\Lambda$CDM)
inflation model.}
\centering
\begin{tabular}{lcccc}
\hline
\hline
\textrm{Data}&
\textrm{TT+lowP ($h\geq0.45$)}&
\textrm{TT+lowP+lensing ($h\geq0.45$)}&
\textrm{TT+lowP+BAO}&
\textrm{TT+lowP+lensing+BAO}\\
\hline
CMB low-$\ell$ & 134.63 (129.83) & 126.26 (126.06) & 155.12 (130.06) & 135.69 (126.00)\\
CMB high-$\ell$ & 151.18 (77.13) & 135.81 (82.46) & 134.19 (73.64)  & 127.58 (80.82)\\
CMB all-$\ell$ & 285.81 (206.97) & 271.84 (218.43) & 293.16 (207.73) & 276.16 (220.31)\\
CMB lensing & --- & 9.77 (9.90) & --- & 10.04 (9.92)\\
BAO & --- & --- & 3.85(4.03) & 2.85 (3.58)\\
\hline
d.o.f.\ & 188 (189) & 196 (197) & 192 (193) & 200 (201) \\
\hline
\hline
\end{tabular}
\end{table*}

It is important to quantitatively understand how well the best-fit 
closed-$\phi$CDM inflation model does relative to the best-fit tilted 
flat-$\Lambda$CDM model in fitting the data. As for the 
closed-$\Lambda$CDM and closed-XCDM cases \citep{Oobaetal2018a, Oobaetal2017}, 
we are unable to resolve this in a quantitative manner, although qualitatively,
overall, the best-fit closed-$\phi$CDM model does not do as well as the 
best-fit tilted flat-$\Lambda$CDM model. However, for some data it does better
than the best-fit closed-$\Lambda$CDM inflation model \citep{Oobaetal2018a}, 
which has one fewer parameter, as well as better than the best-fit closed-XCDM 
case \citep{Oobaetal2017} which has the same number of parameters.

Table \ref{tab:table2} lists the minimum $\chi^2_{\rm eff} = -2 {\rm ln} 
({\rm L_{\rm max}})$ determined from the maximum value of the likelihood, for 
the four data sets we study, for both the closed-$\phi$CDM and tilted
flat-$\Lambda$CDM inflation models, as well as the number of (binned data) 
degrees of freedom (d.o.f.). The d.o.f.\ are determined from combinations of
112 low-$\ell$ TT + lowP, 83 high-$\ell$ TT, 8 lensing CMB (binned) 
measurements, 4 BAO measurements, and 7 (or 6 for the tilted flat-$\Lambda$CDM)
model parameters. Almost certainly the large $\chi^2_{\rm eff}$ values are the 
result of the many nuisance parameters that have been marginalized over, 
as the tilted flat-$\Lambda$CDM model is said to be a good fit to the data.
From this table we see that $\Delta \chi^2_{\rm eff} = 22 (21)$ for the 
closed-$\phi$CDM inflation model (196 (200) d.o.f.),
relative to the tilted flat-$\Lambda$CDM case (197 (201) d.o.f.), 
for the TT + low P + lensing (+ BAO) data 
combination.\footnote{When closed-$\phi$CDM is compared to both 
closed-$\Lambda$CDM and closed-XCDM, $\Delta \chi^2_{\rm eff} = 2 (0)$, 
for these two data sets, where closed-$\Lambda$CDM has the same d.o.f.\ as 
flat-$\Lambda$CDM and closed-XCDM the same as closed-$\phi$CDM 
\citep{Oobaetal2018a, Oobaetal2017}.} 
While this might make the closed-$\phi$CDM model much 
less probable, one can see from the residual panels of Fig. \ref{fig:cls} (e)
\& (f) that this $\Delta \chi^2_{\rm eff}$ is apparently caused by many small 
deviations, and not by a few significant outliers. This allows for the 
possibility that a slight increase in the error bars or a mild non-Gaussianity 
in the errors could raise the model probabilities.    

While there are correlations in the data, it is also instructive to consider 
a standard goodness of fit $\chi^2$ that only makes use of the diagonal 
elements of the correlation matrix. These are listed in Table \ref{tab:table3}
for the four data sets we study and for both the 
closed-$\phi$CDM and tilted flat-$\Lambda$CDM inflation models. From Table
\ref{tab:table3} for the TT + low P + lensing data, we see that the $\chi^2$ 
per d.o.f.\ is 282/196 (228/197, 268/197, 273/196) for the closed-$\phi$CDM 
(tilted flat-$\Lambda$CDM, closed-$\Lambda$CDM, closed-XCDM) inflation model, 
and when BAO data is added to the mix these become 289/200 (234/201, 293/201,
294/200). Again, 
while the closed-$\phi$CDM model is less favored than the tilted 
flat-$\Lambda$CDM case (and is the most favored of the closed models we study 
for the +BAO data combination), it is not straightforward to assess the 
quantitative
significance of this. In addition to the points mentioned at the end of the 
previous paragraph, here we also ignore all the off-diagonal 
information in the correlation matrix, so it is meaningless to compute 
standard probabilities from such $\chi^2$'s. All in all, while the best-fit 
closed-$\phi$CDM inflation model appears less favored, it might be useful 
to perform a more thorough analysis of the model.\footnote{It has previously 
been noted that the constraints derived from higher-$\ell$ and lower-$\ell$ 
Planck 2015 CMB anisotropy data differ slightly 
\citep{Addisonetal2016, Aghanimetal2016}. It might be useful to revisit this
issue using the non-flat models we have studied, to see if this makes a 
difference.}

\section{Conclusion}

We have constrained the physically consistent seven parameter non-flat 
$\phi$CDM inflation model using Planck 2015 CMB data and BAO distance 
measurements.

Unlike the results of the seven parameter non-flat tilted $\Lambda$CDM model 
in \citet{Adeetal2016a}, our seven parameter non-flat $\phi$CDM inflation 
model is not forced to be flat even when the BAO data are added to the mix. 
This is also the case for the non-flat $\Lambda$CDM and non-flat XCDM models we 
studied earlier \citep{Oobaetal2018a, Oobaetal2017, ParkRatra2018a, ParkRatra2018b}.
We find that $\Omega_{\rm k} =  -0.006 \pm 0.005$ at 2$\sigma$ and that 
the best-fit point is about 3$\sigma$ away from flat. In this case the 
improved agreement with the low-$\ell$ $C_{\ell}$ 
observations\footnote{ As noted above, systematics, foreground contamination, and cosmic variance play a bigger roll at lower $\ell$ and we assume that the data error bars correctly account for all effects.} 
 and the weak lensing $\sigma_8$
are not as good compared with the results from the analyses using only the Planck 2015 CMB data.
However, the BAO and CMB data are from very disparate redshifts 
and it is possible that a better model for the intervening epoch or an improved 
understanding of one or both sets of measurements might alter this conclusion.

A more thorough analysis of the non-flat $\Lambda$CDM, XCDM, and $\phi$CDM 
inflation models is needed\footnote{As part of such an analysis, it would be useful to examine smaller-scale predictions of these models, as well as potential CMB spectral distortions \citep[for a review of spectral distortions see][]{Chluba2018}. We note that when fit to the Planck 2015 data non-flat inflation models have a significantly higher reionization optical depth with interesting implications for reionization models \citep{Mitraetal2018}.} to establish if one of them is viable and can 
help resolve some of the low-$\ell$ $C_{\ell}$ issues as well as possibly 
the $\sigma_8$ power issues, without significantly worsening the fit to the 
higher-$\ell$ $C_{\ell}$'s. 
Perhaps non-zero spatial curvature might be more important for 
this purpose than is dark energy dynamics.

\section*{Acknowledgments}

This work is supported by Grants-in-Aid for Scientific Research 
from JSPS (Nos.\ 16J05446 (J.O.) and 15H05890 (N.S.)). B.R.\ is supported 
in part by DOE grant DE-SC0011840.

\pagebreak


\begin{thebibliography}{}

\bibitem[{{Abbott} {et~al.}(2017)}]{Abbottetal2017}
{Abbott}, T.~M.~C. {et~al.} 2017, arXiv:1711.00403

\bibitem[{{Addison} {et~al.}(2016)}]{Addisonetal2016}Addison, G.~E., et~al. 
2016, ApJ, 818, 132 [arXiv:1511.00055]

\bibitem[{{Anderson} {et~al.}(2014)}]{Andersonetal2014}
{Anderson}, L. {et~al.} 2014, MNRAS, 441, 24 [arXiv:1312.4877]

\bibitem[{{Aubourg} {et~al.}(2015)}]{Aubourgetal2015}
{Aubourg}, E. {et~al.} 2015, Phys. Rev. D, 92, 123516 [arXiv:1411.1074]

\bibitem[{{Audren} {et~al.}(2013)}]{Audrenetal2013}
Audren, B,, Lesgourgues, J., Benabed, K., \& Prunet, S. 2013, 
JCAP, 1302, 001 [arXiv:1210.7183]
 
\bibitem[{{Beutler} {et~al.}(2011)}]{Beutleretal2011}
{Beutler}, F. {et~al.} 2011, MNRAS, 416, 3017 [arXiv:1106.3366]

\bibitem[{{Blas} {et~al.}(2011)}]{Blasetal2011}
Blas, D,, Lesgourgues, J., \& Tram, T. 2011, 
JCAP, 1107, 034 [arXiv:1104.2933]

\bibitem[{{Brax} {et~al.}(2000)}]{Braxetal2000}
{Brax}, P, {Martin}, J., \& {Riazuelo}, A. 2000, Phys. Rev. D,  62, 103505 
[arXiv:astro-ph/0005428]

\bibitem[{{Cai} {et~al.}(2016)}]{Caietal2016}
{Cai}, R.-G., {Guo}, Z.-K., \& {Yang}, T. 2016, Phys. Rev. D,  93, 043517 [arXiv:1509.06283]

\bibitem[{{Calabrese} {et~al.}(2012)}]{Calabreseetal2012}
{Calabrese}, E., {Archidiacono}, M., {Melchiorri}, A., \& {Ratra}, B. 2012, Phys. Rev. D,  86, 043520 [arXiv:1205.6753]

\bibitem[{{Chluba}(2018)}]{Chluba2018}
{Chluba}, J. 2018, arXiv:1805.02915

\bibitem[{{Chen} \& {Ratra}(2003)}]{ChenRatra2003}
{Chen}, G., \& {Ratra}, B. 2003, \pasp, 115, 1143 [arXiv:astro-ph/0302002]

\bibitem[{{Chen} \& {Ratra}(2011)}]{ChenRatra2011}
{Chen}, G., \& {Ratra}, B. 2011, \pasp, 123, 1127 [arXiv:1105.5206]

\bibitem[{{Chen} {et~al.}(2017)}]{Chenetal2017}
{Chen}, Y., {Kumar}, S., \& {Ratra}, B. 2017, \apj, 835, 86 [arXiv:1606.07316]

\bibitem[{{Chen} {et~al.}(2016)}]{Chenetal2016}
{Chen}, Y., {et~al.} 2016, \apj, 829, 61 [arXiv:1603.07115]

\bibitem[{Farooq et~al.}(2017)]{Farooqetal2017}
{Farooq}, O., {Madiyar}, F.~R., {Crandall}, S., \& {Ratra}, B. 2017, 
\apj, 835, 26 [arXiv:1607.03537]

\bibitem[{{Farooq} {et~al.}(2015)}]{Farooqetal2015}
Farooq, O., Mania, D., \& Ratra, B. 2015, ApSS, 357, 11 [arXiv:1308.0834]

\bibitem[{Fixsen}(2009)]{Fixsen2009}
Fixsen, D.~J. 2009, \apj, 707, 916 [arXiv:0911.1955]

\bibitem[{{Freedman} {et~al.}(2012)}]{Freedmanetal2012}
{Freedman}, W.~L., {et~al.} 2012, \apj, 758, 24 [arXiv:1208.3281]

\bibitem[{{G{\'o}rski} {et~al.}(1998)}]{Gorskietal1998}
{G{\'o}rski}, K.~M., {et~al.} 1998, \apjs, 114, 1 [arXiv:astro-ph/9608054]

\bibitem[{{G{\'o}rski} {et~al.}(1995)}]{Gorskietal1995}
{G{\'o}rski}, K.~M., {Ratra}, B., {Sugiyama}, N., \& {Banday}, A.~J. 1995,
\apj, 444, L65 [arXiv:astro-ph/9502034]

\bibitem[{{Gott}(1982)}]{Gott1982}
{Gott}, J.~R. 1982, Nature, 295, 304 

\bibitem[{{Hawking} (1984)}]{Hawking1984}
{Hawking}, S.~W. 1984, Nucl. Phys. B, 239, 257

\bibitem[{{Kamionkowski} {et~al.}(1994)}]{Kamionkowskietal1994}
{Kamionkowski}, M., {Ratra}, B., {Spergel}, D.~N., \& {Sugiyama}, N. 1994,
\apj, 434, L1 [arXiv:astro-ph/9406069]

\bibitem[{{Lesgourgues} \& {Tram} (2014)}]{LesgourguesTram2014}
{Lesgourgues}, J., \& {Tram}, T. 2014, JCAP, 1409, 032 
[arXiv:1312.2697]

\bibitem[{{Lewis} {et~al.}(2000)}]{Lewisetal2000}
{Lewis}, A., {Challinor}, A, \& {Lasenby}, A, 2000, \apj, 538, 473 
[arXiv:astro-ph/9911177]

\bibitem[{{L'Huillier} \& {Shafieloo}(2017)}]{LHuillierShafieloo2017}
{L'Huillier}, B., \& {Shafieloo}, A. 2017, JCAP, 1701, 015 [arXiv:1606.06832]

\bibitem[{{Li} {et~al.}(2014)}]{Lietal2014}
{Li}, Y.-L., {Li}, S.-Y., {Zhang}, T.-J., \& {Li}, T.-P. 2014, 
\apj, 789, L15 [arXiv:1404.0773]

\bibitem[{{Li} {et~al.}(2016)}]{Lietal2016}
{Li}, Z., {Wang}, G.-J., {Liao}, K., \& {Zhu}, Z.-H. 2016, 
\apj, 833, 240 [arXiv:1611.00359]

\bibitem[{{Lin} \& {Ishak} (2017)}]{LinIshak2017}
{Lin}, W., \& {Ishak}, M. 2017, arXiv:1708.09813

\bibitem[{{Lukovi{\'c}} {et~al.}(2016)}]{Lukovicetal2016}
{Lukovi{\'c}}, V.~V., {D'Agostino}, R., \& {Vittorio}, N. 2016, 
Astron. Astrophys., 595, A109 [arXiv:1607.05677]
  
\bibitem[{{Mitra} {et~al.}(2018)}]{Mitraetal2018}
{Mitra}, S., {Choudhury}, T.~R., \& {Ratra}, B.\ 2018, \mnras, 479, 4566
[arXiv:1712.00018]

\bibitem[{{Mukherjee} {et~al.}(2003)}]{Mukherjeeetal2003}
{Mukherjee}, P., {et~al.} 2003, \apj, 598, 767 [arXiv:astro-ph/0306147]

\bibitem[{{Ooba} {et~al.}(2018a)}]{Oobaetal2018a}
{Ooba}, J., {Ratra}, B., \& {Sugiyama}, N. 2018, \apj, in press 
[arXiv:1707.03452]

\bibitem[{{Ooba} {et~al.}(2017)}]{Oobaetal2017}
{Ooba}, J., {Ratra}, B., \& {Sugiyama}, N. 2017, arXiv:1710.03271

\bibitem[{{Ooba} {et~al.}(2018b)}]{Oobaetal2018b}
{Ooba}, J., {Ratra}, B., \& {Sugiyama}, N.\ 2018b, arXiv:1802.05571

\bibitem[{{Park} \& {Ratra}(2018a)}]{ParkRatra2018a}
{Park}, C.-G., \& {Ratra}, B.\ 2018a, arXiv:1801.00213

\bibitem[{{Park} \& {Ratra}(2018b)}]{ParkRatra2018b}
{Park}, C.-G., \& {Ratra}, B.\ 2018b, arXiv:1803.05522

\bibitem[{{Park} \& {Ratra}(2018c)}]{ParkRatra2018c}
{Park}, C.-G., \& {Ratra}, B.\ 2018c, arXiv:1807.07421

\bibitem[Pavlov {et~al.}(2013)]{Pavlovetal2013}
{Pavlov}, A., Westmoreland, S., Saaidi, K., \& Ratra, B. 2013, Phys. Rev. D, 88, 123513 [arXiv:1307.7399]

\bibitem[{{Peebles}(1984)}]{Peebles1984}
{Peebles}, P.~J.~E. 1984, \apj, 284, 439

\bibitem[{{Peebles} \& {Ratra}(1988)}]{PeeblesRatra1988}
{Peebles}, P.~J.~E., \& {Ratra}, B. 1988, \apj, 325, L17

\bibitem[{{Planck Collaboration}(2016a)}]{Adeetal2016a}
Planck Collaboration 2016a, Astron.\ Astrophys., 594, A13 [arXiv:1502.01589]

\bibitem[{{Planck Collaboration}(2016b)}]{Aghanimetal2016}
Planck Collaboration 2016b, arXiv:1608.02487

\bibitem[{{Podariu} \& {Ratra}(2001)}]{PodariuRatra2001}
{Podariu}, S., \& {Ratra}, B. 2001, \apj, 532, 109
[arXiv:astro-ph/9910527]

\bibitem[{{Rana} {et~al.}(2017)}]{Ranaetal2017}
{Rana}, A., {Jain}, D., {Mahajan}, S., \& {Mukherjee}, A. 2017, 
JCAP, 1703, 028 [arXiv:1611.07196]

\bibitem[{{Ratra}(1985)}]{Ratra1985}
{Ratra}, B. 1985, Phys. Rev. D, 31, 1931

\bibitem[{{Ratra}(2017)}]{Ratra2017}
{Ratra}, B. 2017, Phys. Rev. D, 96, 103534 [arXiv:1707.03439] 

\bibitem[{{Ratra} \& {Peebles}(1988)}]{RatraPeebles1988}
{Ratra}, B., \& {Peebles}, P.~J.~E. 1988, Phys. Rev. D, 37, 3406

\bibitem[{{Ratra} \& {Peebles}(1994)}]{RatraPeebles1994}
{Ratra}, B., \& {Peebles}, P.~J.~E. 1994, \apj, 432, L5

\bibitem[{{Ratra} \& {Peebles}(1995)}]{RatraPeebles1995}
{Ratra}, B., \& {Peebles}, P.~J.~E. 1995, Phys. Rev. D, 52, 1837

\bibitem[{{Riess} {et~al.}(2016)}]{Riessetal2016}
{Riess}, A.~G., {et~al.} 2016, \apj, 826, 56 [arXiv:1604.01424]

\bibitem[{{Ross} {et~al.}(2015)}]{Rossetal2015}
{Ross}, A.~J. {et~al.} 2015, MNRAS, 449, 835 [arXiv:1409.3242]

\bibitem[{{Ryan} {et~al.}(2018)}]{Ryanetal2018}
{Ryan}, J., {Doshi}, S., \& {Ratra}, B.\ 2018, \mnras, 480, 759 
[arXiv:1805.06408]

\bibitem[Sapone {et~al.}(2014)]{Saponeetal2014}
{Sapone}, D., Majerotto, E., \& Nesseris, S. 2014, 
Phys. Rev. D, 90, 023012 [arXiv:1402.2236]

\bibitem[{{Sievers} {et~al.}(2013)}]{Sieversetal2013}
{Sievers}, J.~L., {et~al.} 2013, JCAP, 1310, 060 [arXiv:1301.0824]

\bibitem[{{Sol\`{a}} {et~al.}(2018)}]{Solaetal2018}
{Sol\`{a}}, J., {de Cruz P\'{e}rez}, J., \& {G\'{o}mez-Valent}, A.\ 2018, 
Europhys.\ Lett., 121, 39001 [arXiv:1606.00450]

\bibitem[{{Sol\`{a}} {et~al.}(2017c)}]{Solaetal2017c}
{Sol\`{a}}, J., {de Cruz P\'{e}rez}, J \& {G\'{o}mez-Valent}, A. 2017c, 
arXiv:1703.08218

\bibitem[{{Sol\`{a}} {et~al.}(2017a)}]{Solaetal2017a}
{Sol\`{a}}, J., {G\'{o}mez-Valent}, A., \& {de Cruz P\'{e}rez}, J. 2017a, 
\apj, 836, 43 [arXiv:1602.02103] 

\bibitem[{{Sol\`{a}} {et~al.}(2017b)}]{Solaetal2017b}
{Sol\`{a}}, J., {G\'{o}mez-Valent}, A., \& {de Cruz P\'{e}rez}, J. 2017b, 
Mod. Phys. Lett. A, 32, 1750054 [arXiv:1610.08965]

\bibitem[{{Starobinsky}(1996)}]{Starobinsky1996}
{Starobinsky}, A.~A. 1996, arXiv:astro-ph/9603075

\bibitem[{{Wang} {et~al.}(2017)}]{Wangetal2017}
Wang, Y., Xu, L., \& Zhao, G.-B. 2017, arXiv:1706.09149

\bibitem[{{Wei} \& {Wu}(2017)}]{WeiWu2017}
Wei, J.-J., \& Wu, X.-F. 2017, \apj, 838, 160 [arXiv:1611.00904]

\bibitem[{{White} \& {Scott}(1996)}]{WhiteScott1996}
White, M., \& Scott, D. 1996, \apj, 459, 415 [arXiv:astro-ph/9508157]

\bibitem[{{Yu} {et~al.}(2018)}]{Yuetal2018}
{Yu}, H., {Ratra}, B., \& {Wang}, F.-Y.\ 2018, \apj, 856, 3 [arXiv:1711.03437]

\bibitem[{{Yu} \& {Wang}(2016)}]{YuWang2016}
{Yu}, H., \& {Wang}, F.~Y. 2016, \apj, 828, 85 [arXiv:1605.02483]

\bibitem[{{Zaldarriaga} {et~al.}(1998)}]{Zaldarriagaetal1998}
{Zaldarriaga}, M., Seljak, U., \& {Bertschinger}, E. 1998, \apj, 494, 491 
[arXiv:astro-ph/9704265] 

\bibitem[{{Zhang} {et~al.}(2017)}]{Zhangetal2017}
{Zhang}, X., {et~al.} 2017, Res. Astron. Astrophys., 17, 6 [arXiv:1703.08293]

\end{thebibliography}
\end{document}